# Pressure Dependent Structural Changes and Predicted Electrical Polarization in Perovskite $RMnO_3$

T. Wu[1], T. A. Tyson[1, 4], H. Chen[1], P. Gao[1], T. Yu[1], Z. Chen[2], Z. Liu[3],

K. H. Ahn[1,4] , X. Wang[4] and S.-W. Cheong[4]

[1] Department of Physics, New Jersey Institute of Technology, Newark, NJ 07102
[2] Mineral Physics Institute, Stony Brook University, Stony Brook, NY 11794
[3] Geophysical Laboratory, Carnegie Institution of Washington, DC 20015
[4] Rutgers Center for Emergent Materials and Department of Physics & Astronomy, Rutgers University, Piscataway, NJ 08854

Corresponding Author:
Trevor A. Tyson, E-mail: tyson@njit.edu

## Abstract

High pressure x-ray diffraction (XRD) measurements on *R*$MnO_3$ (*R*=Dy, Ho and Lu) reveals that varying structural changes occurs for different *R* ions. Large lattice changes (orthorhombic strain) occur in $DyMnO_3$ and $HoMnO_3$ while the Jahn-Teller (JT) distortion remains stable. On the other hand, in $LuMnO_3$, Mn-O bond distortions are observed in the region 4-8 GPa with the broad minimum in the JT distortion. High pressure IR measurements indicate that a phonon near 390 $cm^{-1}$ corresponding to the complex motion of the Mn and O ions changes anomalously for $LuMnO_3$. It softens in the 4-8 GPa region, which is consistent with the structural change in Mn-O bonds and then hardens at high pressures. By contrast, the phonons continuously harden with increasing pressure for $DyMnO_3$ and $HoMnO_3$. DFT calculations show that the *E*-phase $LuMnO_3$ is the most stable phase up to the 10 GPa pressure examined. Simulations indicate that the distinct structural change under pressure in $LuMnO_3$ can possibly be used to optimize the electric polarization by pressure/strain.

## I. Introduction

The manganties ($RMnO_3$) have attracted much attention and have been extensively studied for both basic physics and applications perspectives since they exhibit exceptionally interesting properties, especially the cross coupling between the magnetism and ferroelectricity (multiferroic behavior)[1]. The coexistence of these two order parameters brings out multiferroics as a promising class of multifunctional materials. For example, the power consumption of the magnetic random-access memories (MRAMs) is expected to be lowered by reversing the spin states with electric field[2]. Moreover, smaller device size and high density data storage can be achieved with multiple functions integrated into one material.

Two crystal phases, hexagonal and orthorhombic, exist for $RMnO_3$ at ambient pressure[3]. The orthorhombic structure (*Pbnm*) is stable for large *R* ions (La, Pr, Nd, Tb and Dy), while $RMnO_3$ with small *R* ions (Ho, Er, Yb, Lu, In and Sc) adopts the hexagonal structure ($P6_3cm$, at room temperature). However, orthorhombic $RMnO_3$ with small *R* ion, such as Ho and Lu, can be synthesized as a metastable perovskites by high pressure synthesis technique[4] or special chemical methods[5]. Perovskite $RMnO_3$ consists of corner sharing $MnO_6$ octahedra, with the Mn ions in the center of the octahedral. The *R* ions are at the center of a cube formed by eight $MnO_6$ octahedra (see Fig. 1). This structure is highly distorted because of the mismatch of *R*-O and Mn-O bond distances and the Jahn-Teller (JT) distortion of the Mn-O bonds.

The discovery of multiferroicity (coupled magnetization and electrical polarization) in $TbMnO_3$ and $DyMnO_3$ has stimulated significant interests in investigating the magnetic, ferrolectric and structural properties on the perovskite manganites in recent years[6, 7]. Hence, it is important to understand the magnetoelectric mechanism for the perovskite manganites. The magnetic phase diagram is strongly dependent on the structural details of these materials[8, 9]. Large *R* ions such as La exhibit *A*-type spin configuration, with ferromagnetic spin order in the *a*-*b* plane and stacked antiferromagnetically along the *c* axis, with the magnetic ordering temperature $T_N$ ~ 140 K. $T_N$ monotonically decreases with decreasing *R* ions size (or decrease of the Mn-O-Mn bond angles since the rotation of the $MnO_6$ octahedra depends strongly on the size of *R* ions). The transverse spiral antiferromagnetic structure becomes the stable ground state for the smaller *R* ions Tb and Dy. This group of non-collinear spin ordering can generate ferroelectricity by the Dzyaloshinskii-Moriya interaction[10, 11]. The spiral spin order breaks the spatial inversion symmetry and ferroelectricity and magnetoelectric effect are allowed. The spin order changes again for the even smaller *R* ion size (Y to Sc) into the *E*-type antiferromagnetic phase. The nature of ferroelectricity in *E*-phase materials is currently under detailed study sparked by work by Sergineko et al[13] who estimate that E-phase systems would have electric polarization values 2 orders of magnitude larger than the TbMnO3 system[12],[13]. Sergienko *et al.* proposed the double-exchange interaction model between the Mn *d* orbitals in the *E*-phase. In follow on work, the electric polarization (*P*) for orthorhombic $HoMnO_3$ was estimated to be ~ 6 $\mu$C cm$^{-2}$ by *ab initio* density functional theory (DFT) calculations[14].

Much experimental and theoretical work has been done starting from the initial studies of E-phase $RMnO_3$[14-21]. However, the full expected value of electric polarization has yet to be achieved. Electric polarization measurements by the positive-up negative-down (PUND) method suggested $P$ ~ 0.24 $\mu$C·cm$^{-2}$ for the value extrapolated to 0 K for single crystal $HoMnO_3$[21]. Recently, $P$ was measured to be 0.15 $\mu C \cdot cm^{-2}$ at 2 K along $c$ axis in single crystal[19]. Perovskite $LuMnO_3$ was also well studied including the crystal structure and ferroelectric properties[9, 17, 22-24]. $P$ measurements by the PUND method on perovskite $RMnO_3$ ($R$=Ho, Tm, Yb and Lu) polycrystalline samples shown a continuously increasing value with decreasing $R$ ion size. Thus, $LuMnO_3$ has the largest polarization in this series with $P$=0.17 $\mu C \cdot cm^{-2}$ at 6 K. The estimated maximum $P$ for single crystal $LuMnO_3$ is ~0.6 $\mu C \cdot cm^{-2}$ at 0 K[25]. The value of $P$ in $E$-phase *Lu*$MnO_3$ was found to be ~ 0.4 $\mu C \cdot cm^{-2}$ at 2 K by Ishiwata *et al*[26].

Perovskike $RMnO_3$ has been extensively investigated by varying external parameters, such as temperature, magnetic field, chemical doping and strains (thin film). However, not much work has been conducted for the perovskite $RMnO_3$ as a function of high pressure, especially for the small $R$ ion $RMnO_3$. Pressure dependent measurements give a unique perspective to explore the crystal structure, electron and spin correlations in an well controlled manner. An example of the application of external pressure is the high pressure structural study of $LaMnO_3$ which probed the JT distortion, orbital order and insulator-metal transition[27]. It should also be noted that high pressure research on multiferroic $RMnO_3$, including $TbMnO_3$[28], $DyMnO_3$[29] and $GdMnO_3$[30] has been published recently. Chen *et al.* (Ref. 28) performed high pressure X-ray diffraction (XRD) and X-ray absorption spectroscopy (XAS) as well as *ab initio* calculations on both $TbMnO_3$ and $DyMnO_3$. It was reported that the JT distortion was suppressed and the bandwidth was broadened with hydrostatic pressure. High pressure (up to 53 GPa) has been applied to $GdMnO_3$ for the Raman and XRD measurements. It was suggested that a structural phase transition occurred at 50 GPa with a space group change from *Pnma* to $P2_13$. For *E*-phase $RMnO_3$, the in-plane strain effects on the ferroelectricity and magnetism for $HoMnO_3$ studied by DFT calculations and a model Hamiltonian technique[20]. It was shown that the electric polarization was significantly enhanced under the compressive strain, due to the increase of the electronic contribution. However, the E-phase becomes unstable. The asymmetric hopping of $e_g$ electrons corresponding to the orbital order change under the strain induces a large value of $P$.

In this work, XRD measurements under hydrostatic pressure were conducted on the *E*-phase $HoMnO_3$ and $LuMnO_3$ and compared to $DyMnO_3$. IR measurements under pressure and DFT simulations on the pressure dependence of the electric polarization are conducted to complements the XRD measurements. The crystal structure and phonon response was studied in detail. It was found that $LuMnO_3$ has a different compression behavior compared with $DyMnO_3$ and $HoMnO_3$. A broad minimum in the JT distortion occurs in the region between 4 and 8 GPa. The phonons associated with the complex motion of the Mn and O ions changes anomalously in the same pressure range. Density functional theory (DFT) methods were applied to predict the low temperature magnetic order and electric polarization as a function of pressure. *E*-phase $LuMnO_3$ is the most stable structure. This result suggests that the electrical polarization in $LuMnO_3$ system can be optimized by the pressure/strain and it is promising in the sensor applications.

## II. Experimental and Modeling Methods

Polycrystalline hexagonal $R$MnO$_3$ ($R$=Ho and Lu) and orthorhombic DyMnO$_3$ were prepared by solid state reaction. Synthesis of orthorhombic HoMnO$_3$ and LuMnO$_3$ starting from the hexagonal phase was carried out at the High Pressure Laboratory of the Mineral Physics Institute, Stony Brook University, using a 2000-ton split-sphere multi-anvil apparatus (USSA-2000). A 14/8 cell assembly was used, which consists of eight WC cubes (25 mm) with the 8 mm truncations as the second stage anvils, a ceramic MgO octahedron with the edge-length of 14 mm as pressure medium, a graphite sleeve was used as the resistive furnace, and a gold capsule as the sample container. The cell assembly with hexagonal phase sample inside was cold-compressed to the target oil pressure of 6 GPa. Then it was heated to 1300 °C followed by quenching to room temperature and single phase orthorhombic HoMnO$_3$ and LuMnO$_3$ were obtained.

The perovskite $R$MnO$_3$ ($R$=Dy, Ho and Lu) were prepared by grinding and sieving the powder (500 mesh) for the synchrotron based high pressure x-ray diffraction (XRD) measurements up to ~ 25 GPa with a diamond anvil cell (DAC) at the beamline X17C at the National Synchrotron Light Source (NSLS), in Brookhaven National Laboratory (BNL). Methanol/Ethanol/Water (16:3:1) mixture was used as the pressure medium to generate hydrostatic pressure. XRD images were collected with a Rayonix 165 charge coupled device (CCD) detector with a focused monochromic X-ray beam $\lambda$=0.4066 Å. The program Fit2D was utilized to integrate the two dimensional diffraction image to yield the one dimensional intensity versus 2$\theta$ XRD pattern[31]. Rietveld refinements on the XRD data were conducted by the program TOPAS academic. The space group *Pbnm* (#62) was adopted for refinements. Pressure dependent infrared (IR) absorption spectra were conducted up to ~ 20 GPa at the beam line U2A at NSLS. A vacuum infrared microscope provided resolution of 4 $cm^{-1}$ in far IR (150 - 700 $cm^{-1}$) range. It was equipped with a 3.5-micron Mylar beam splitter, a 40 mm working distant reflecting objective and a Si bolometer detector. CsI was used to dilute the samples and acted as the pressure medium in IR measurements.

DFT calculation were conducted on $LuMnO_3$ and $HoMnO_3$. The approach adapted by Picozzi *et al* for ambient pressure and strained films was utilized in the DFT simulations (see Ref. 20 and references therein). All DFT structural optimization simulations were conducted using 40 atoms cells to reflect the magnetic order for the *E*-phase. Simulations utilized the experimentally derived lattice parameters but the atomic positions were optimized to reduce the forces on the atoms to less than 0.003 eV/Å. Total energies at each pressure were thus found for each spin configuration (FM, *A*-AFM and *E*-AFM) and the FM configuration was use as a reference. At each pressure the *E*-phase electric polarization was computed (ionic and electronic). The electronic contribution to the polarization utilized the berry phase method. To understand the phonon modes of these system the phonon density of states was computed (for $HoMnO_3$) from the force constants obtained from frozen phonon simulation for a 2 x 2 x 2 cell (160 atoms). The calculation of the phonon density of states follow the methods of Ref. [43(a)] implemented in the Phonopy codes [43(b)].

## III. Results and Discussion

Figure 2(a), 2(b) and 2(c) show the XRD patterns for $R$MnO$_3$ ($R$=Dy, Ho and Lu) at selected pressures, respectively. All the samples have similar patterns since they are all in orthorhombic structure with the same space group *Pbnm* and similar lattice parameters. The diffraction peaks shift to high 2θ angle, drop in intensity and broaden the width with increasing pressure. No new peak appears in the whole pressure range for all the samples indicating a continuous compression process on the samples without phase change.

Rietveld refinements were performed on the XRD data to obtain the structural parameters. The profiles of the refinements for $R$MnO$_3$ ($R$=Dy, Ho and Lu) at 1 GPa are shown in the Figure 3(a), 3(b) and 3(c). A mask(excluded region as dark band) was applied for the extra peak from the steel gasket in HoMnO$_3$ (Figure 3(b)). The observed (crosses), calculated (solid line) and difference (bottom line) profiles are shown. The vertical bars display the peak positions of the reflections of the model structure. The weighted profile $R$ factors ($R_{wp}$) for $R$MnO$_3$ (R=Dy, Ho and Lu) were 0.0834, 0.0806 and 0.0703, respectively. Lattice and atomic parameters were obtained for $R$MnO$_3$ (R=Dy, Ho and Lu) up to 11 GPa and only lattice parameters were available for the complete pressure range.

Figure 4(a), 4(b) and 4(c) give the lattice compressibility for $R$MnO$_3$ ($R$=Dy, Ho and Lu) as a function of pressure. (Note that a weak bump appears near 14 GPa for all the samples resulting from the glass transition of the pressure medium.) DyMnO$_3$ and HoMnO$_3$ have the same compression behavior for the lattice parameters since Dy and Ho

are close to each other in the periodic table with similar ion radius. The *a* and *b* parameters are compressed at a similar rate with increasing pressure, while *b* is softer than *a* at low pressure. The order reverses above ~ 18 GPa. *c* is the least compressible direction. For LuMnO3 with smaller radius, the compression becomes more isotropic. No inversion of the order occurs in *a* and *b* compression. *c* decreases in similar slope as *a* and *b*.

The pressure dependent volume for $R\text{MnO}_3$ ($R$=Dy, Ho and Lu) is shown in Figure 4(d), 4(e) and 4(f). A first order equation of state fit by the Murnaghan equation[32] was performed. The data points near 14 GPa were not included. The bulk modulus $B_0$ were obtained for $R\text{MnO}_3$ ($R$=Dy, Ho and Lu) with the values, 117.61±6.33 GPa (Dy), 117.95±7.04 GPa (Ho) and 104.03±3.89 GPa (Lu), respectively. These values are consistent with the typical perovskite $\text{LaMnO}_3$ which is 108±2 GPa[27]. It indicates that $\text{DyMnO}_3$ and $\text{HoMnO}_3$ have the same compressibility and they are stiffer than $\text{LuMnO}_3$. Interstingly, LuMnO3 is closer to the well-known $\text{LaMnO}_3$ system. The softness of $\text{LuMnO}_3$ can be attributed to the small *R* ion. The large value of $B_0^{'}$ (pressure derivative of bulk modulus) for these samples indicates significant anisotropic compression (typical $B_0^{'}$ values for crystals with isotropic compression are in the range 4-6[27]).

The distortion in the unit cell at high pressures is further described by the orthorhombic strains in the *ab* plane and along the *c* axis which are defined as $Os_{ab}=2(c-a)/(c+a)$ and $O_c=2(a+b-2^{1/2}c)/(a+b+2^{1/2}c)$, respectively[33]. Figure 5 gives the orthorhombic strains for all the samples as a function of pressure. $\text{DyMnO}_3$ (square symbols) and $\text{HoMnO}_3$ (circle symbols) change with the same trend in both $Os_{ab}$ and $O_c$ (see Figure 5(a) and 5(b)). When the pressure is below 10 GPa, $Os_{ab}$ and $O_c$ slightly converge to 0 which indicates the tendency to become cubic like. Then $Os_{ab}$ and $O_c$ drop with increasing pressure up to 25 GPa. Both $Os_{ab}$ and $O_c$ in $LuMnO_3$ (triangle symbols) decrease generally in the whole pressure range (Figure 5(a) and 5(b)). The weak slope indicates the small pressure effect on the lattice in $LuMnO_3$ compared to $DyMnO_3$ and $HoMnO_3$. The bump in Figure 5(b) is from the glass transition of the pressure medium as mentioned before.

We focus on the region below ~10 GPa. Figures 6(a), 6(b) and 6(c) give the Mn-O bond distances as a function of pressure for $RMnO_3$ ($R$=Dy, Ho and Lu). Mn-O1($s$), Mn-O2($m$) and Mn-O2($l$) denote the short apical Mn-O bond, the medium and long equatorial Mn-O bonds in the $MnO_6$ octahedron, respectively. No obvious change occurs in Mn-O bonds for $DyMnO_3$ and $HoMnO_3$ with increasing pressure (see Figure 6(a) and 6(b)). Mn-O2($l$) bonds maintain values of ~ 2.2 Å and ~ 2.1 Å for $DyMnO_3$ and $HoMnO_3$, respectively. Slight decrease of Mn-O1($s$), Mn-O2($m$) bonds is observed at high pressures above ~ 8 GPa. Note that the difference between long and short Mn-O bonds in $DyMnO_3$ is larger than that in $HoMnO_3$. It indicates a more symmetric $MnO_6$ octahedra in $HoMnO_3$ in this pressure range.

In the case of LuMnO3, the pressure effect is much stronger on Mn-O bonds (see Figure 6(c)). The Mn-O2($l$) bond decreases from 2.2 Å to 2.1 Å with increasing pressure to 4 GPa and keeps constant up to 11 GPa. Mn-O1($s$) and Mn-O2($m$) bonds increase from ~ 1.9 Å to ~ 2.0 Å at low pressure. A broad maximum occurs in the pressure range 4-8GPa following with a significant decrease at high pressure. The overall effect is that there is an intermediate region of pressure where the $MnO_6$ polyhedra become less distorted with regions of high distortion on either side. Hence, pressure has the largest effect on the $MnO_6$ octahedra with the biggest Mn-O bond change in $LuMnO_3$ (compared to R=Dy and Ho) while $LuMnO_3$ has the most stable unit cell shape as shown in Figure 5.

To parameterize the octahedral distortion, Fig. 7 shows the pressure dependence of $\Delta_d$ which is defined as $(1/6)\sum_{n=1,6}[((\text{Mn-O})_n - \langle\text{Mn-O}\rangle)/\langle\text{Mn-O}\rangle]^2$ for $R$$MnO_3$ ($R$=Dy, Ho and Lu). $\Delta_d$ measures the deviation of the Mn-O bond distances from the mean value in $MnO_6$ octahedron and it gives the magnitude of the JT distortion. In $DyMnO_3$ and $HoMnO_3$, JT distortion is stable with pressure change, while the distortion in $DyMnO_3$ is larger. However, $\Delta_d$ changes significantly as a function of pressure in $LuMnO_3$ which shows a U shape. A broad minimum is reached in the 4-8 GPa region which is the same pressure range as the broad maximum in the Mn-O1(*s*) and Mn-O2(*m*) bond distances (see Figure 6(c)).

Thus, it was found $DyMnO_3$ and $HoMnO_3$ have the similar structural change with increasing pressure, while $LuMnO_3$ is compressed in different behavior. Significant lattice parameter change occurs in $DyMnO_3$ and $HoMnO_3$, while Mn-O bonds are stable resulting in the constant JT distortion. Weak effects for the lattice change are observed in $LuMnO_3$. However, Mn-O bonds and JT distortion change significantly as a function of pressure. The broad maximum of the Mn-O1(*s*) and Mn-O2(*m*) bond distances causes the minimum JT distortion. This result is consistent with the previous theories which suggested the $MnO_6$ octahedral rotation (lattice change) can stabilize the JT distortion[34].

To complement and support structural measurements under pressure, IR measurements under the pressure were performed. Figure 8(a), 8(b) and 8(c) show the IR optical density (OD) spectra for $R$$MnO_3$ ($R$=Dy, Ho and Lu) as a function of pressure.

All the samples have the similar profiles since and no very large changes occurs with increasing pressures up to 16 GPa. The inset in each panel shows the details of the spectra in the range 370-490 $cm^{-1}$. Three solid lines in the inset show the positions of the three component peaks. For $DyMnO_3$ and $HoMnO_3$, all three peaks shift to high frequency indicating the hardening phonons with pressure as expected. However, an anomaly of the phonon change with pressure occurs in $LuMnO_3$. The phonon at ~ 390 $cm^{-1}$ softens at low pressures and then hardens at high pressures.

To understand the phonon shift as a function of pressure in detail, the three peaks in 370-490 $cm^{-1}$ were fit and Figure 9 shows the pressure dependent phonon frequency for $R$MnO$_3$ ($R$=Dy, Ho and Lu). The inset in Figure 9(b) gives the fit details. The data points and the fitting curve overlap well indicating the high fit quality. The three lower curves in the inset show the three component peaks. In Figure 9(a), all three phonons for $DyMnO_3$ and $HoMnO_3$ harden with increasing pressure. The phonons at ~ 425 $cm^{-1}$ and 450 $cm^{-1}$ in $LuMnO_3$ increase monotonically with pressure (see Figure 9(b)). However, the phonon at ~390 $cm^{-1}$ slightly softens at low pressures following with hardening. A broad minimum in ~4-8 GPa is in the same pressure range as the structural change in the $MnO_6$ octahedra. Identification of these phonon mode will enable connection with the structural data.

Extensive studies have been conducted on the IR phonons of $AMnO_3$ type perovskite systems[35-41]. Hence, we complement these results with our DFT projected phonon density of states calculation in Fig. 10 where we show that the intermediate region near 400 $cm^{-1}$ has is dominated by O oxygen contributions and Mn contribution with limited contribution from the Ho or (R sites). Our ab initio calculations are consistent with recent results from semi-empirical shell models (R=Tb, Dy and Ho)[42]. We note that the IR spectra are proportional to the phonon density times the transition matrix element between to initial states of the system according to Fermi's Golden Rule. Hence, Mn and O ion motions are involved in the anomalous phonon mode. The specific phonon mode was identified by DFT frozen phonon calculations. The previous work indicates that there are three distinct regions corresponding to collective motions of $B_{1u}$, $B_{2u}$ and $B_{3u}$ symmetry of atoms in the structures.

Symmetry analysis of the modes reveals details of the atomic motions. The modes near 400 $cm^{-1}$ correspond to complex motion of the Mn and O ions with very small displacements of the A ions but large and similar displacement amplitudes Mn/O ions. These modes are sensitive to the Mn-O-Mn bond angles. The DFT simulations of the phonon modes show the motion of unique atoms for this mode (see Figure 11). At lower energies, in finer details, it was noted in previous work that the tilting, buckling and rotation of $MnO_6$ octahedra cover the range ~300 to 500 $cm^{-1}$ while bond stretching in internal modes within the $MnO_6$ polyhedra cover the range 500 to 600 $cm^{-1}$. [43]

Again in Figure 11 we see one of the calculated modes for $HoMnO_3$ near 400 $cm^{-1}$ showing that they alter the Mn-O-Mn bond angles. It is noted that the Mn and Ho sites are all equivalent in the unit cell so the displacements of the Mn and O ions does indeed change the Mn-O-Mn bond angles. We recall for the experimental IR data that, while the $DyMnO_3$ and $HoMnO_3$ systems exhibit continuous hardening of the phonon frequencies in the IR peak positions as a function of pressure, phonons in $LuMnO_3$ soften and then harden with increasing pressure in the region between 4 and 8 GPa. The IR results are then consistent with the minimum in the distortion of the Mn-O bonds found in 4-8 GPa in XRD measurements for $LuMnO_3$. We note that the measurements here are for the high temperature phase with no magnetic order. However, we expect that sensitivity to pressure should carry over to the low temperature and the polarization characteristics of the E phase polarization although the exact details of the trends may be different. The results suggest possible use of $LuMnO_3$ system in strain/pressure sensors. The electrical polarization can be optimized by the pressure/strain.

Density functional theory (DFT) calculations were applied for $HoMnO_3$ and $LuMnO_3$ in order to predict the stable magnetic structure and ferroelectric polarization in the low temperature magnetically ordered phase. Figure 12(a) shows the energy of *A*-type and *E*-type antiferromagnetic structure relative to the ferromagnetic state for $HoMnO_3$ and $LuMnO_3$. *E*-phase $LuMnO_3$ has the lowest energy and it becomes even more stable with increasing pressure. This is contrast to the case of strained films where in plane compression destabilized the *E*-phase[20]. Figure 12(b) gives the electric polarization ($P$), including the ionic and electronic contributions, as a function of pressure for *E*-phase $HoMnO_3$ and $LuMnO_3$ expected for the low temperature behavior in the magnetically ordered phase. $P$ increases with increasing pressure for both samples and it reaches 6.5 $\mu C/cm^2$ and 5.7 $\mu C/cm^2$ at 10 GPa for $HoMnO_3$ and $LuMnO_3$, respectively. However, *E*-phase $LuMnO_3$ is the most stable phase.

## IV. Summary

In summary, high pressure XRD measurements on $R$$MnO_3$ ($R$=Dy, Ho and Lu) reveal that varying structural change as a function of pressure occurs for different $R$ ions.

$DyMnO_3$ and $HoMnO_3$ have similar compression behavior, although $DyMnO_3$ is more distorted than $HoMnO_3$. Large lattice change (orthorhombic strain) occurs in $DyMnO_3$ and $HoMnO_3$ under the pressure, while Mn-O bonds in $MnO_6$ octahedra are stable and no obvious JT distortion change is observed. Significant Mn-O bonds change occurs in $LuMnO_3$. Thus, large JT distortions are observed with a broad minimum between 4 and 8 GPa. High pressure IR measurements show that the phonon at ~390 $cm^{-1}$ corresponding to the complex motion of the Mn and O ions changes anomalously in $LuMnO_3$. It softens in the region 4-8 GPa following with hardening at higher pressures. This result is consistent with the structural change in the $MnO_6$ octahedra which has the lowest distortion between 4 GPa and 8 GPa. The phonons in $DyMnO_3$ and $HoMnO_3$ continuously harden with increasing pressure. DFT calculations are used to predict behavior in the low temperature magnetically ordered phase.

## V. Acknowledgments

Work at NJIT is supported by DOE Grant DE-FG02−07ER46402 (TY, TAT, HC, PG,TY) and at Rutgers University by DOE Grant DE-FG02−07ER46382 (XW and SWC). Synchrotron powder X-ray diffraction was performed at Brookhaven National Laboratory's National Synchrotron Light Source (NSLS) which is funded by the U.S. Department of Energy.

## Figure Captions

**Figure 1.** Crystal structure of orthorhombic $R$MnO$_3$ showing the MnO$_6$ octahedra with significant distortion.

**Figure 2.** High pressure synchrotron XRD patterns for $R$MnO$_3$ ($R$=Dy, Ho, and Lu) at selected pressures in (a), (b) and (c), respectively.

**Figure 3.** Profiles of Rietveld refinements for $R$MnO$_3$ ($R$=Dy, Ho and Lu) at 1 GPa in (a), (b) and (c), respectively. The observed (crosses), calculated (solid line) and difference (bottom line) profiles are shown. The vertical bars show the peak positions of reflections. Mask was applied in (b) to exclude the extra peak from the steel gasket.

**Figure 4.** Pressure dependent compressibility ($\Delta a/a_0$, $\Delta b/b_0$ and $\Delta c/c_0$) in (a), (b) and (c) and volume in (d), (e) and (f) for $R$MnO$_3$ ($R$=Dy, Ho and Lu), respectively. The solid lines in panels (d), (e) and (f) show the first order equation of state fits by Murnaghan equation.

**Figure 5.** The orthorhombic strains in *ab* plane ($Os_{ab}$) and along *c* axis ($Os_c$) as a function of pressure for $R$MnO$_3$ ($R$=Dy, Ho and Lu) in (a) and (b), respectively.

**Figure 6.** Pressure dependence of Mn-O1 apical and Mn-O2 equatorial bonds in $MnO_6$ octhedra for $RMnO_3$ ($R$=Dy, Ho and Lu) in (a), (b) and (c), respectively. Note that broad maximum of Mn-O1 (*s*) and Mn-O2 (*m*) reaches in 4 - 8 GPa for $LuMnO_3$.

**Figure 7.** $\Delta_d$ as a function of pressure for $RMnO_3$ ($R$=Dy, Ho and Lu) showing the deviation of the Mn-O bond distances from the mean value. Note that broad minimum of $\Delta_d$ occurs in 4 – 8 GPa for $LuMnO_3$.

**Figure 8.** Pressure dependent infrared absorption spectra of $RMnO_3$ ($R$=Dy, Ho and Lu) in (a), (b) and (c), respectively. The insets in each panel show the magnification of the phonons in the range 370 - 490 $cm^{-1}$ for all the samples. The solid lines in the insets display the positions of the three component peaks as a function of pressure.

**Figure 9.** Pressure dependence of phonon frequencies between 370 $cm^{-1}$ and 490 $cm^{-1}$ for $RMnO_3$ ($R$=Dy and Ho) in (a) and $LuMnO_3$ in (b). Note that abrupt change of phonon at ~ 390 $cm^{-1}$ occurs in $LuMnO_3$ near 6 GPa which is in the same pressure range as the structural anomaly.

**Figure 10.** Computed atomic site project phonon density of states for HoMnO3 for a 2 x 2 x 2 $HoMnO_3$ cell (160 atoms) revealing contributions from the Ho, Mn, and O sites. Note that the contribution of Ho is important only at low frequencies.

**Figure 11.** Phonon mode in $HoMnO_3$ in the intermediate region near 400 $cm^{-1}$ corresponding to collective motion of Mn and O ions which change the Mn-O-Mn bond angles. Small, medium and large spheres correspond to the O, Mn and Ho atoms. The magnitude of the ion displacement is proportional to the length of the vectors.

**Figure 12.** (a) Energy of *A*-type and *E*-type antiferromagnetic structure relative to the ferromagnetic state as a function of pressure for $RMnO_3$ ($R$=Ho and Lu). Note that *E*-type $LuMnO_3$ is the most stable phase with the lowest energy. (b) Pressure dependent total electric polarization for $RMnO_3$ ($R$=Ho and Lu). The results correspond to behabior in the low temperature magnetically ordered state.

Fig. 1. Wu *et al*.

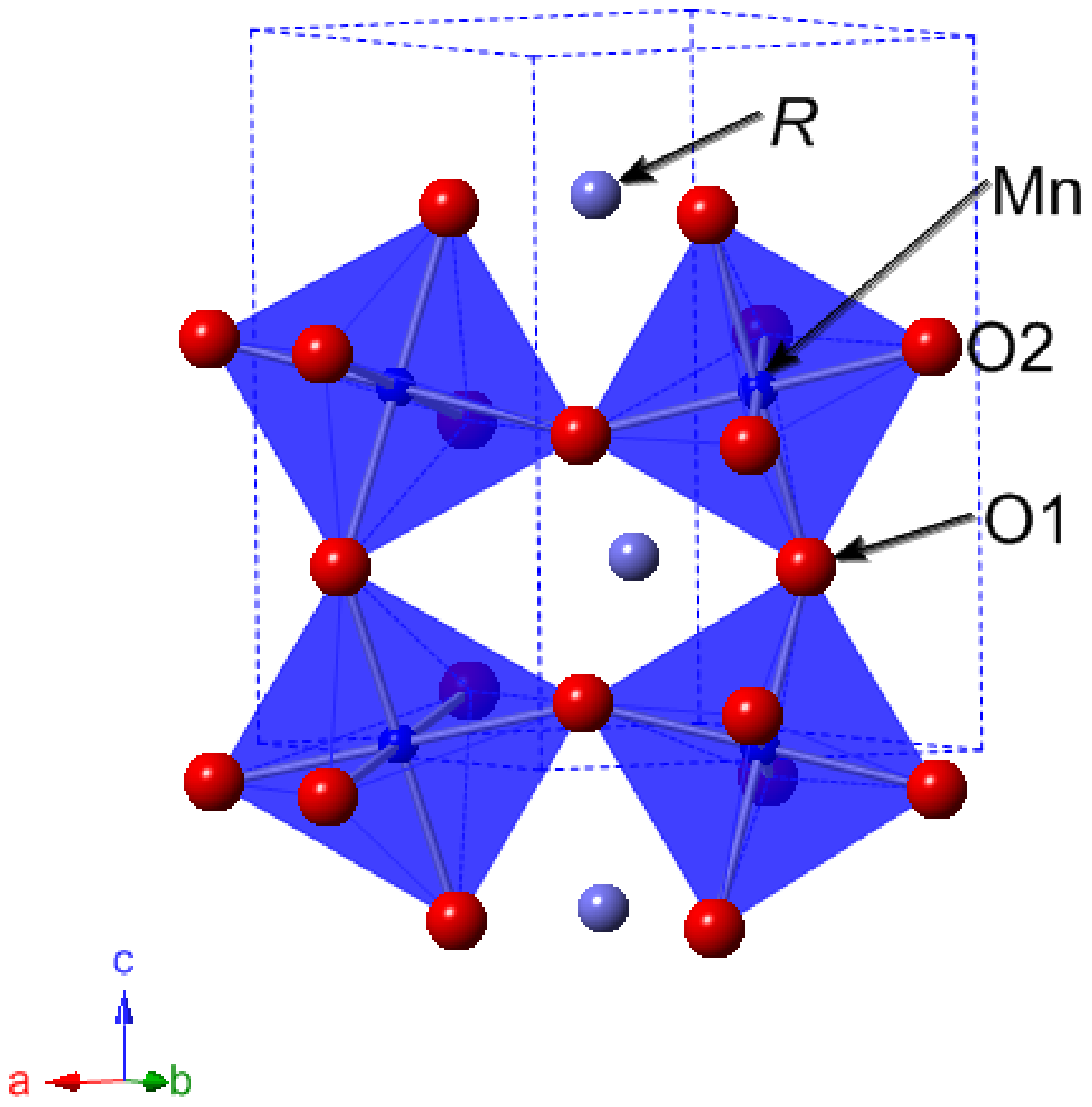

Fig. 2. Wu *et al*.

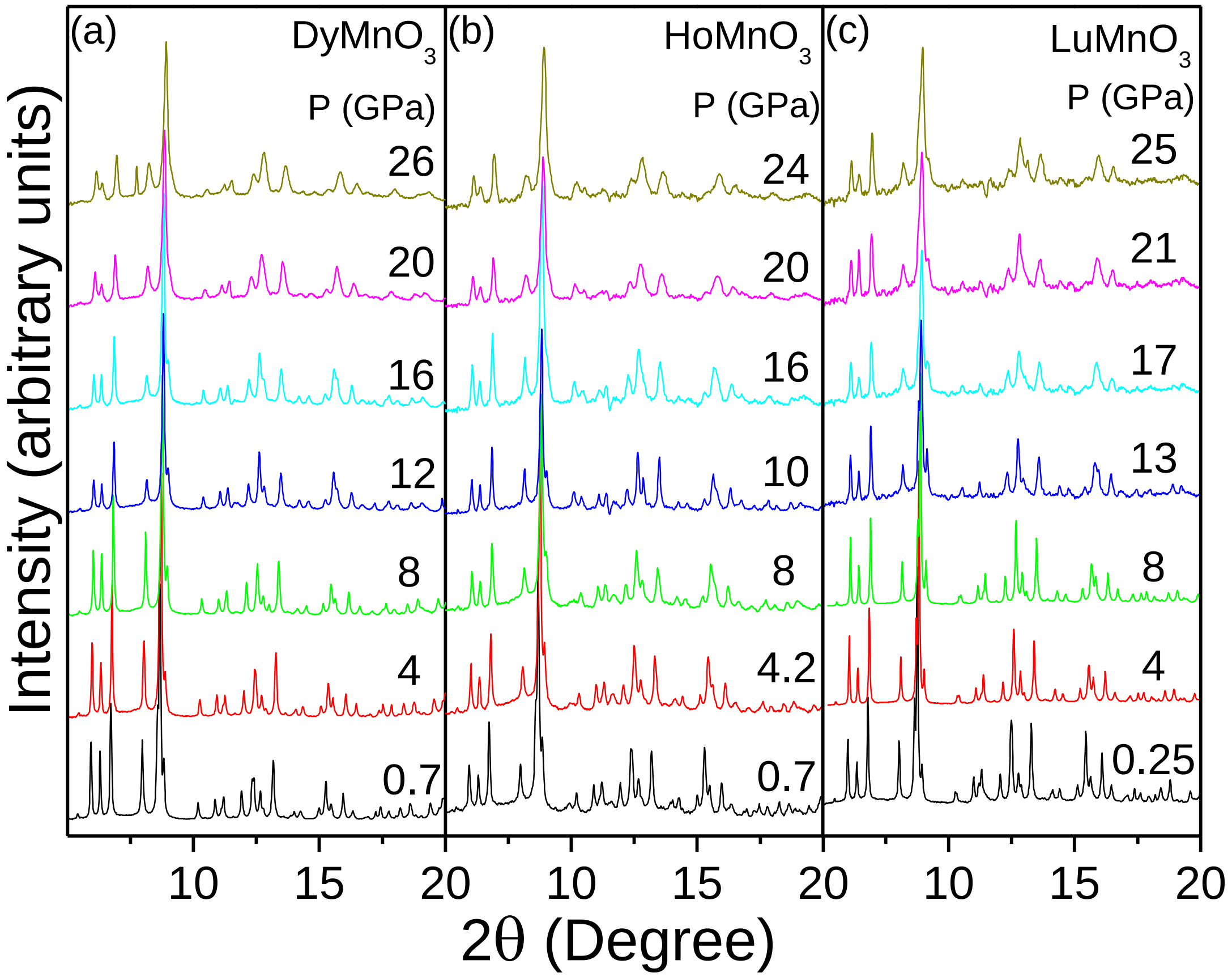

Fig. 3. Wu *et al*.

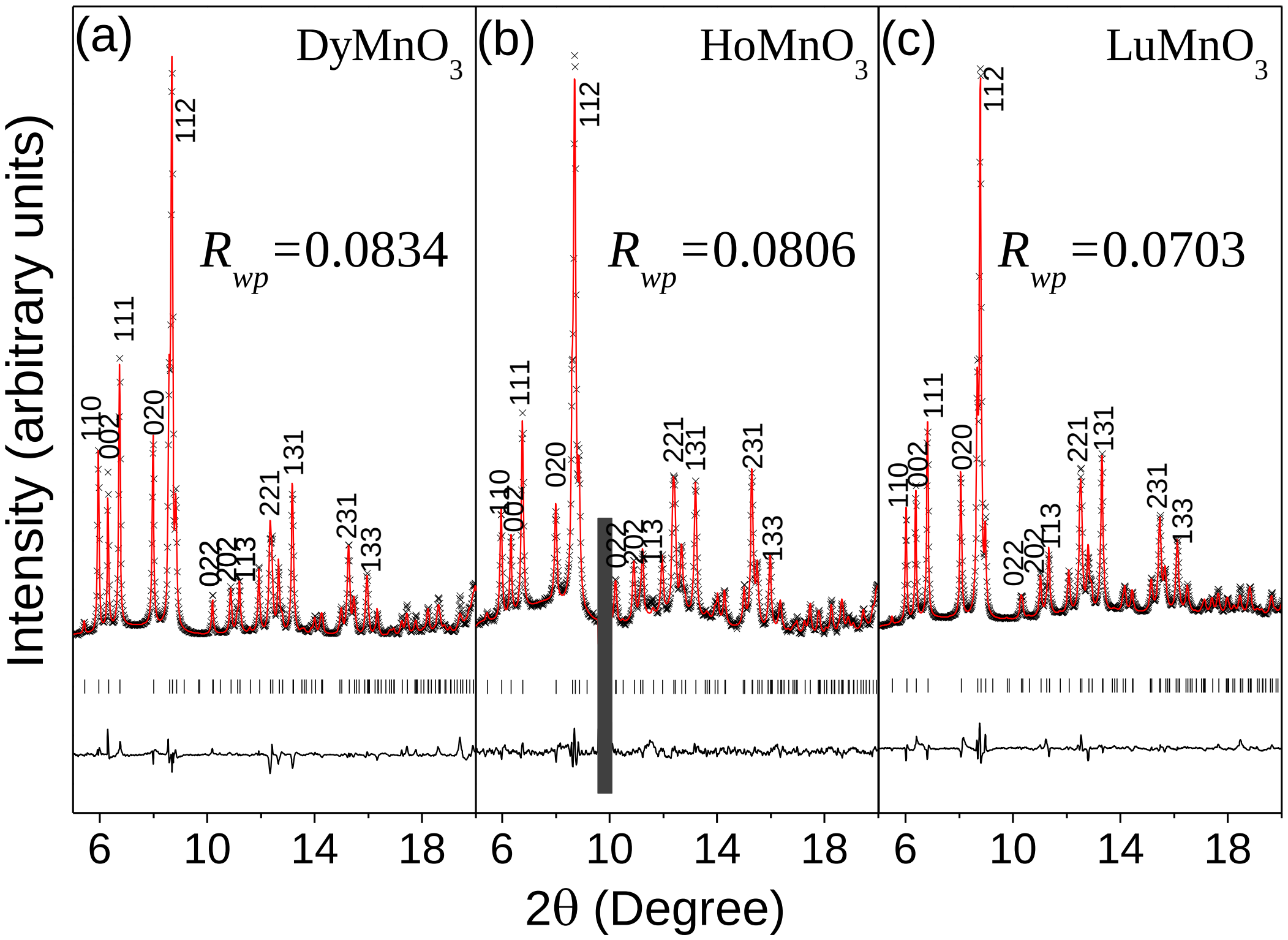

Fig. 4. Wu *et al*.

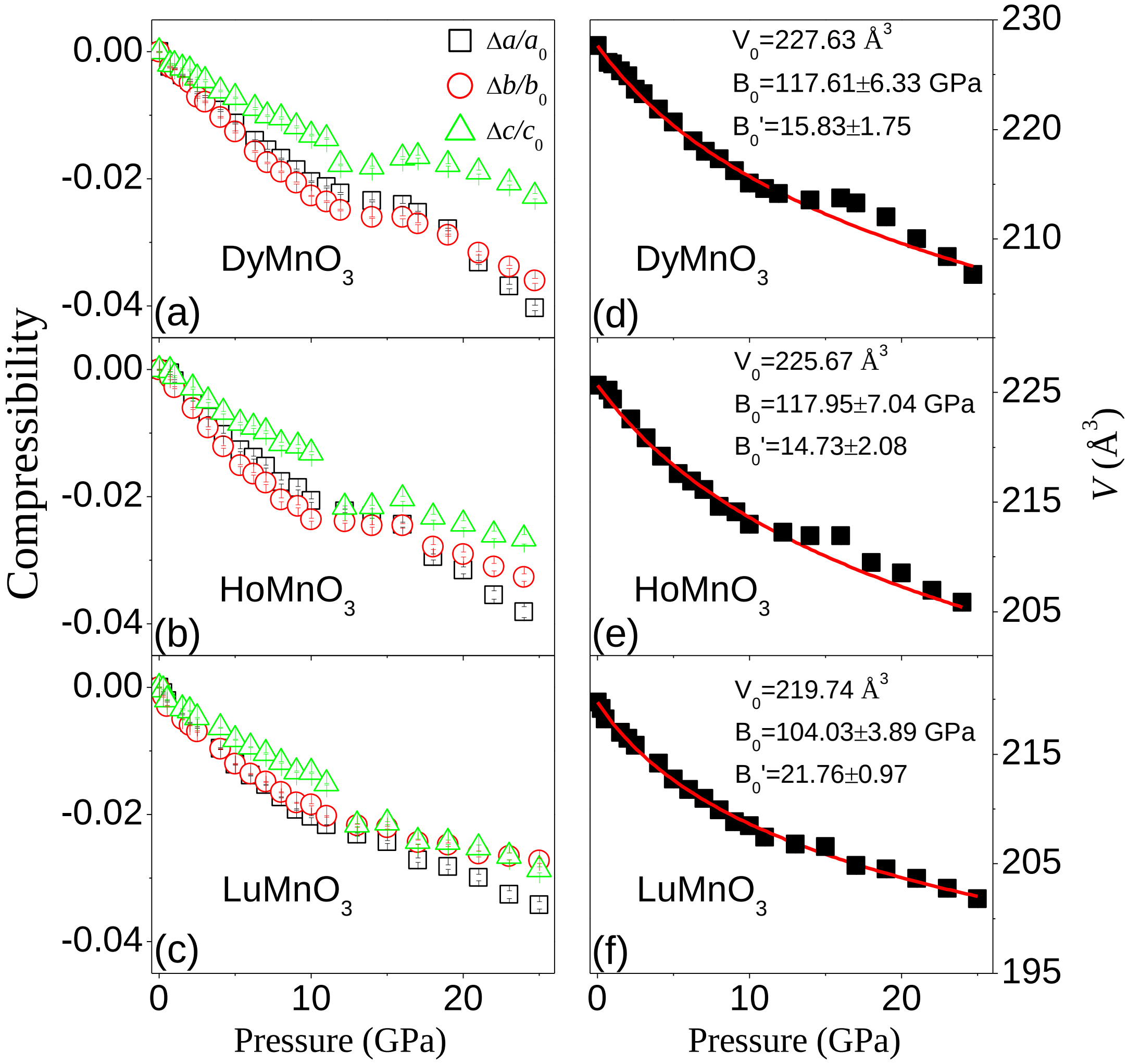

Fig. 5. Wu *et al*.

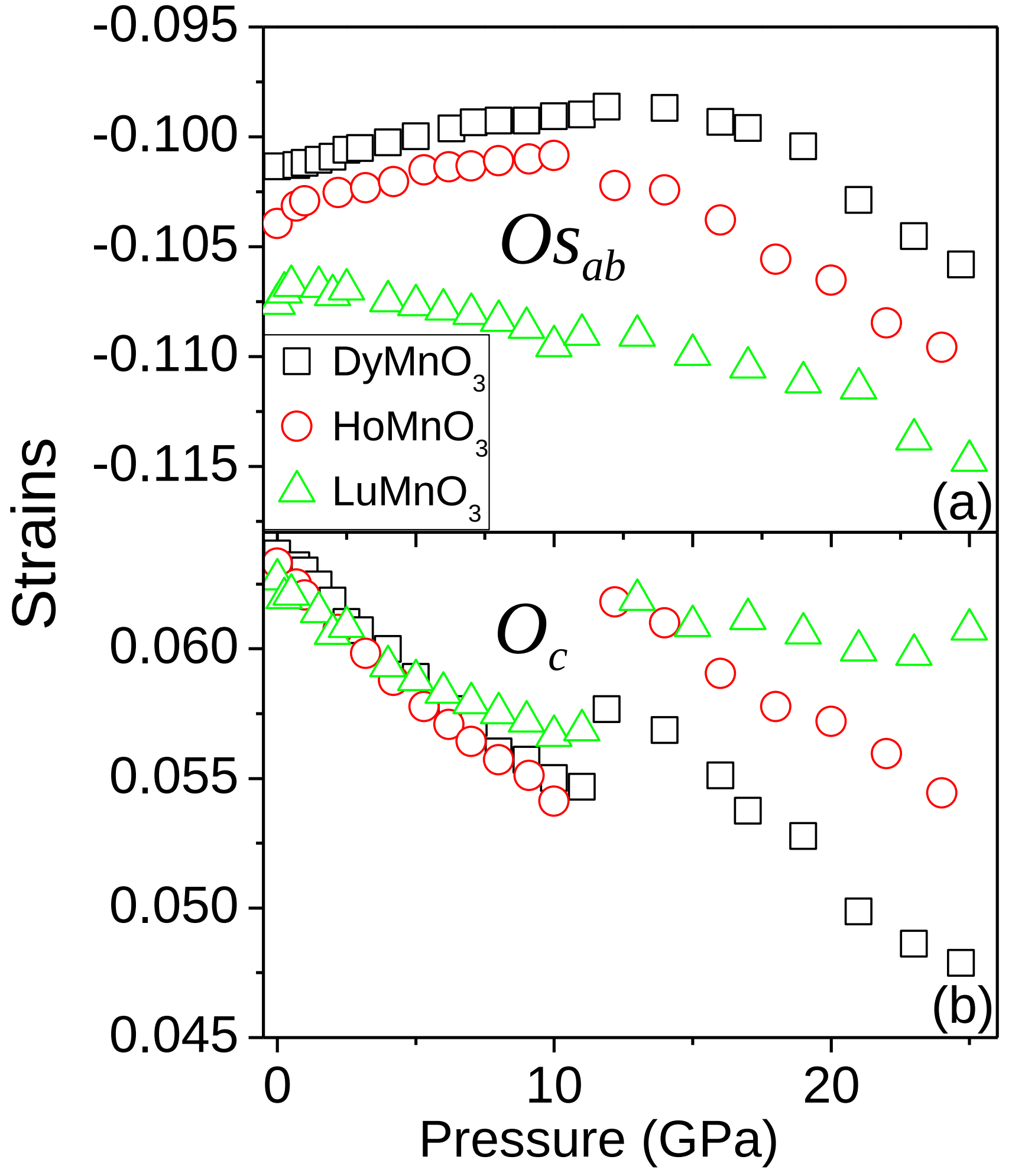

Fig. 6. Wu *et al*.

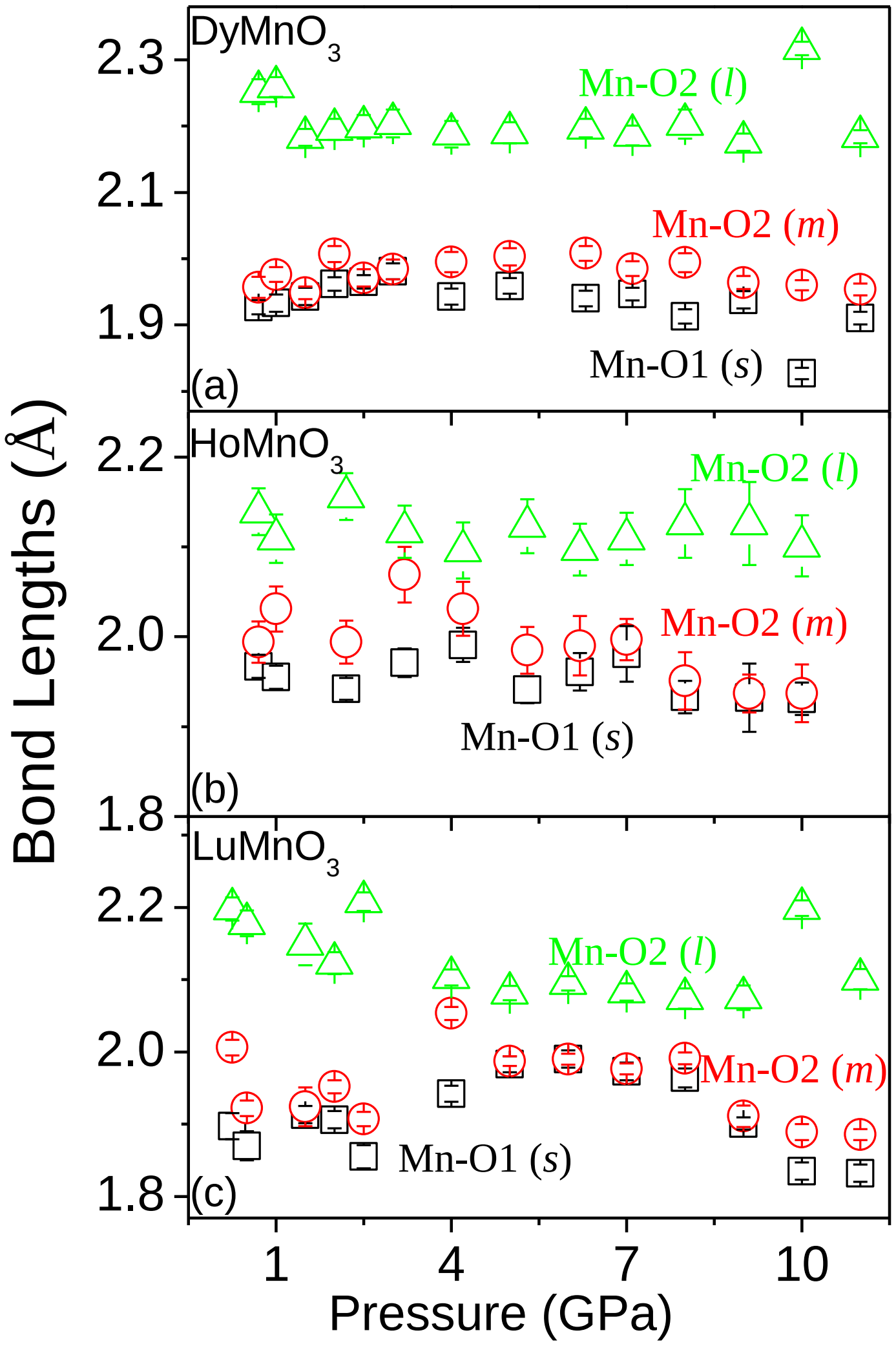

Fig. 7. Wu *et al*.

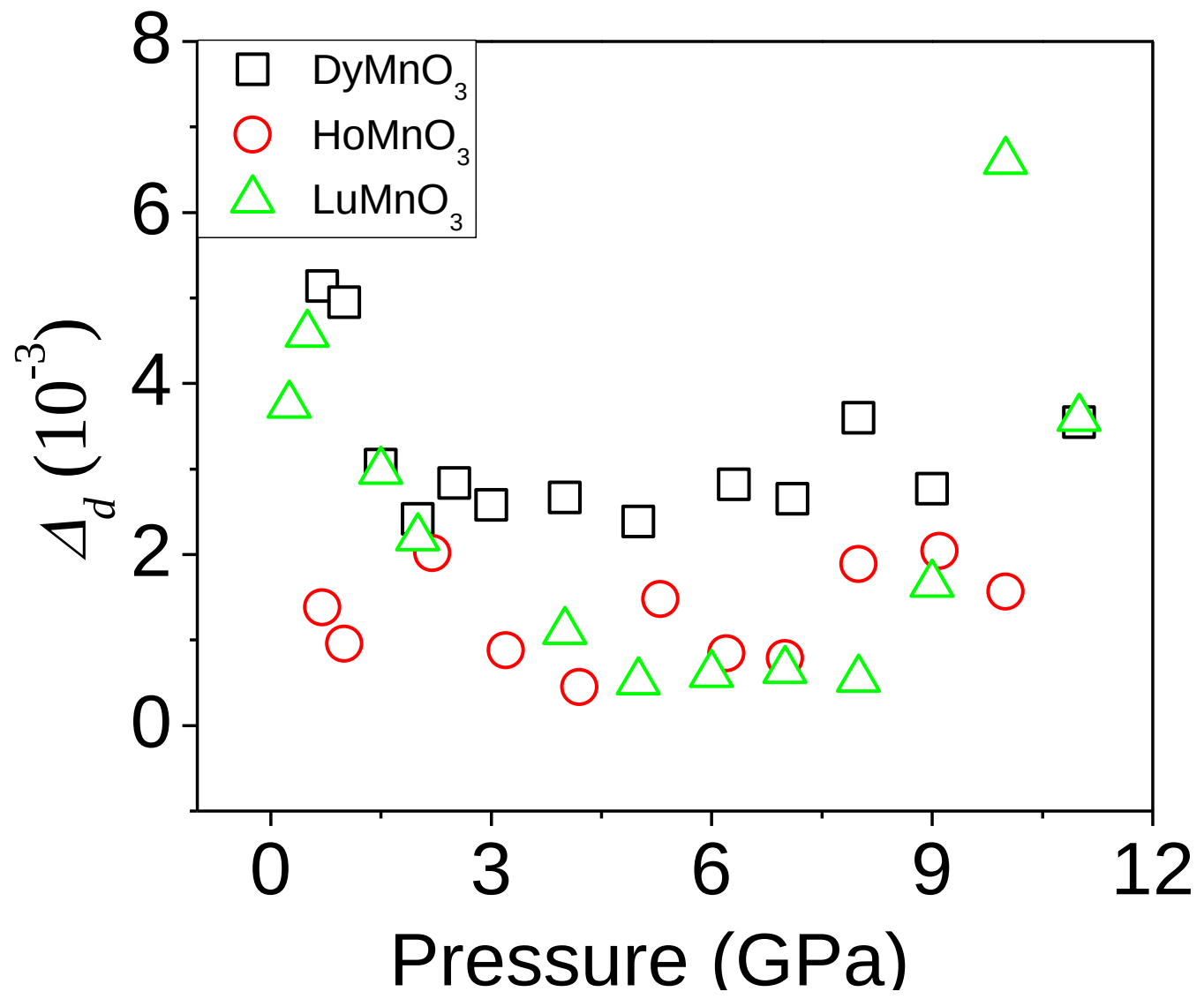

Fig. 8. Wu *et al*.

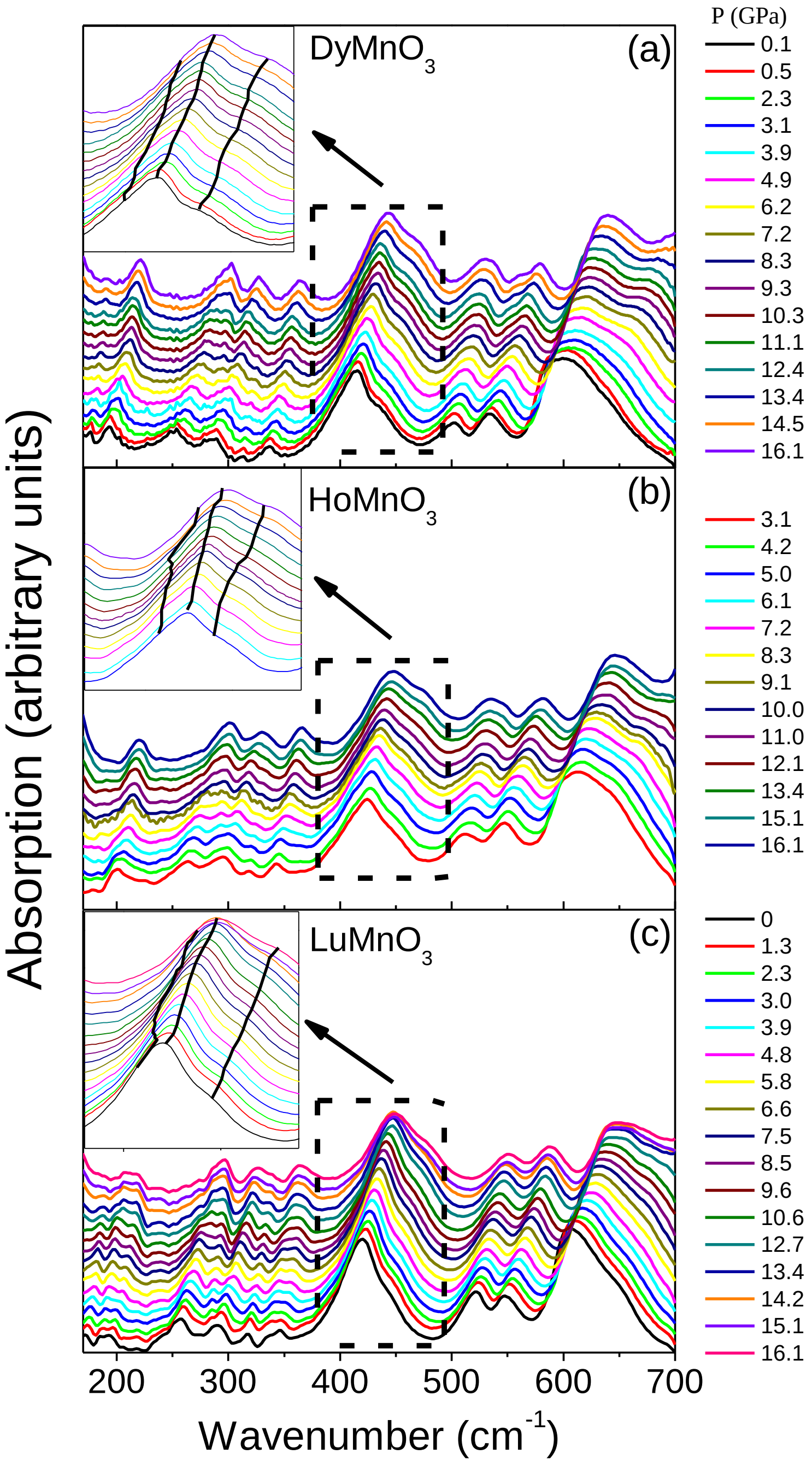

Fig. 9. Wu *et al*.

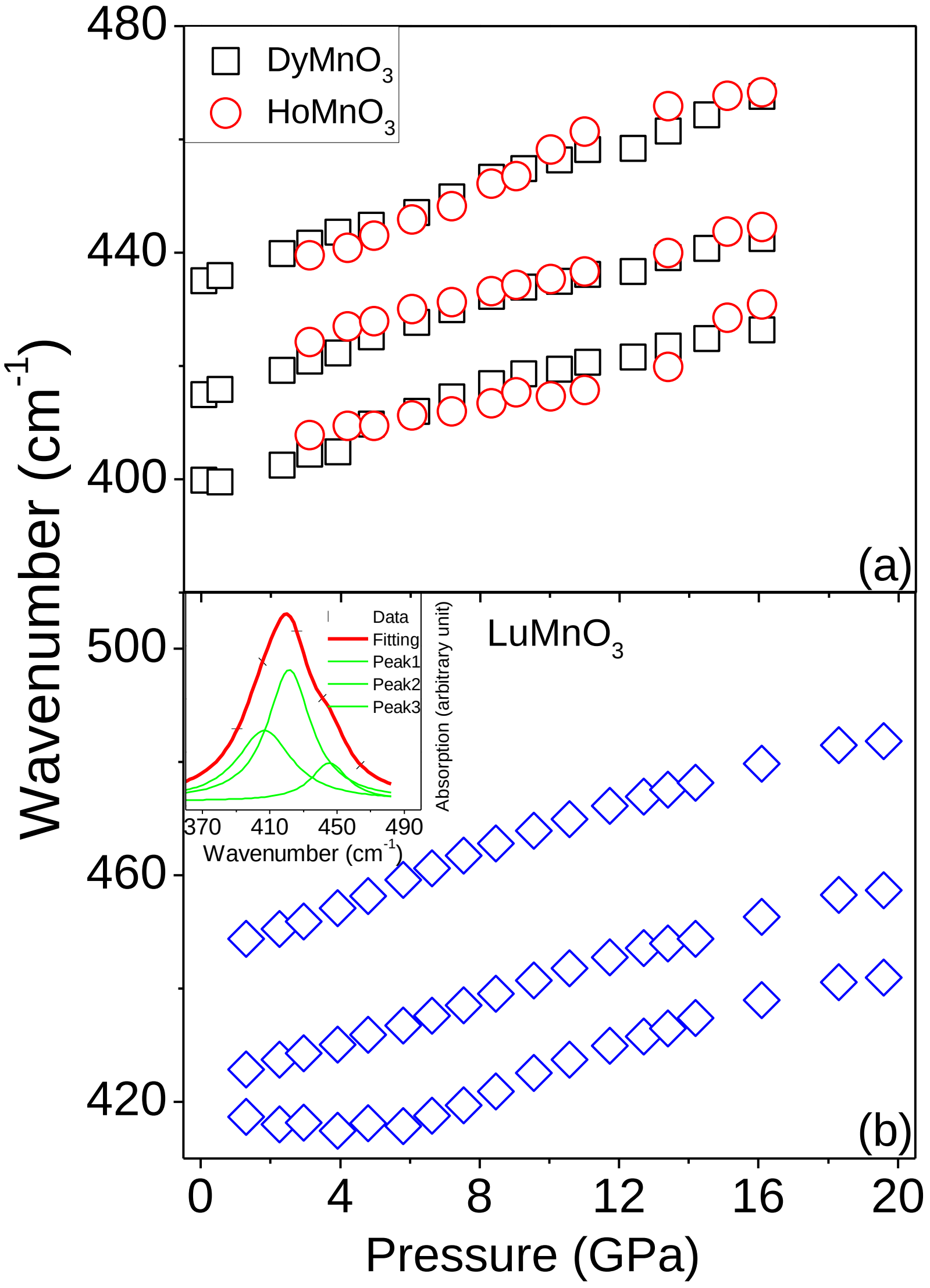

Fig. 10. Wu *et al*.

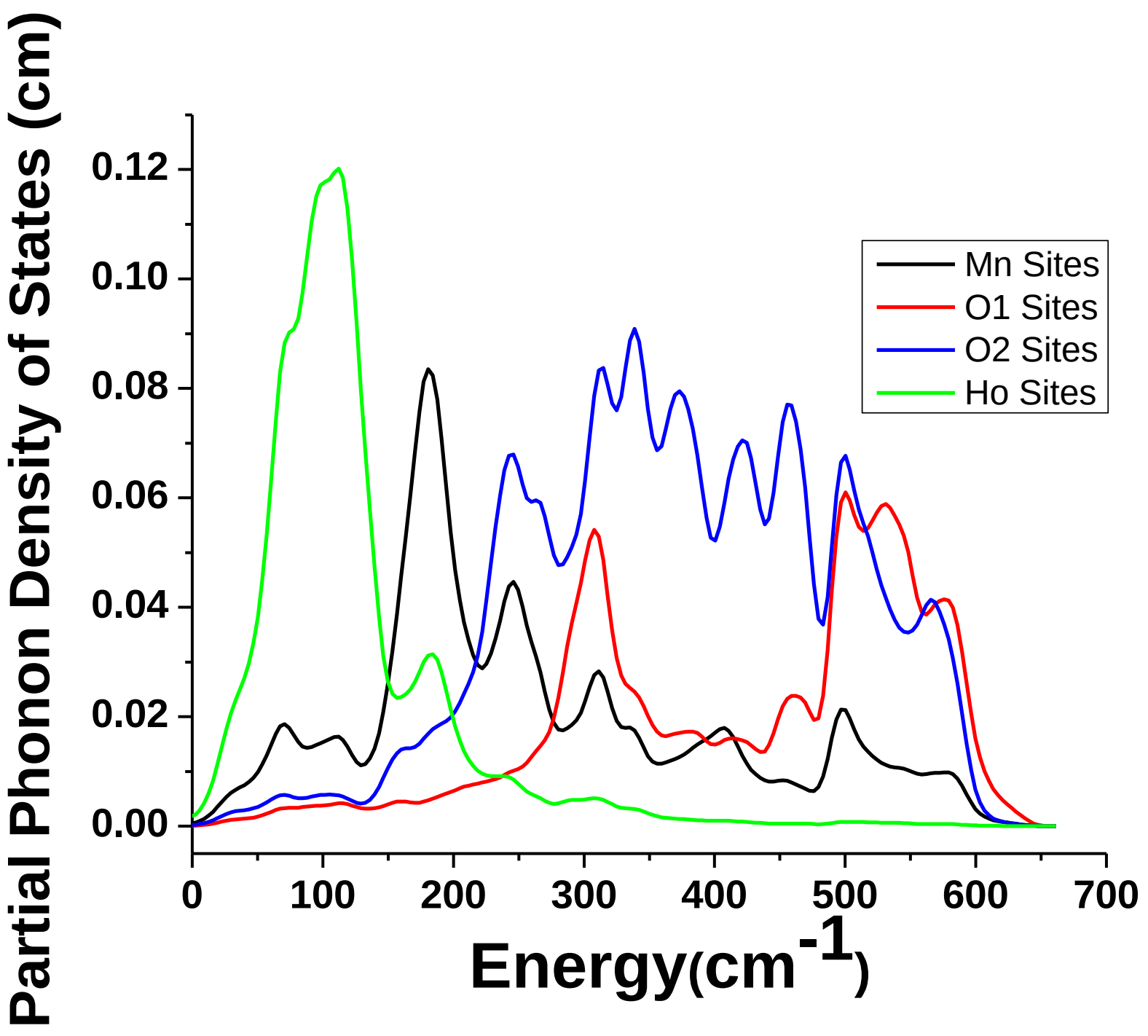

Fig. 11.  Wu *et al*.

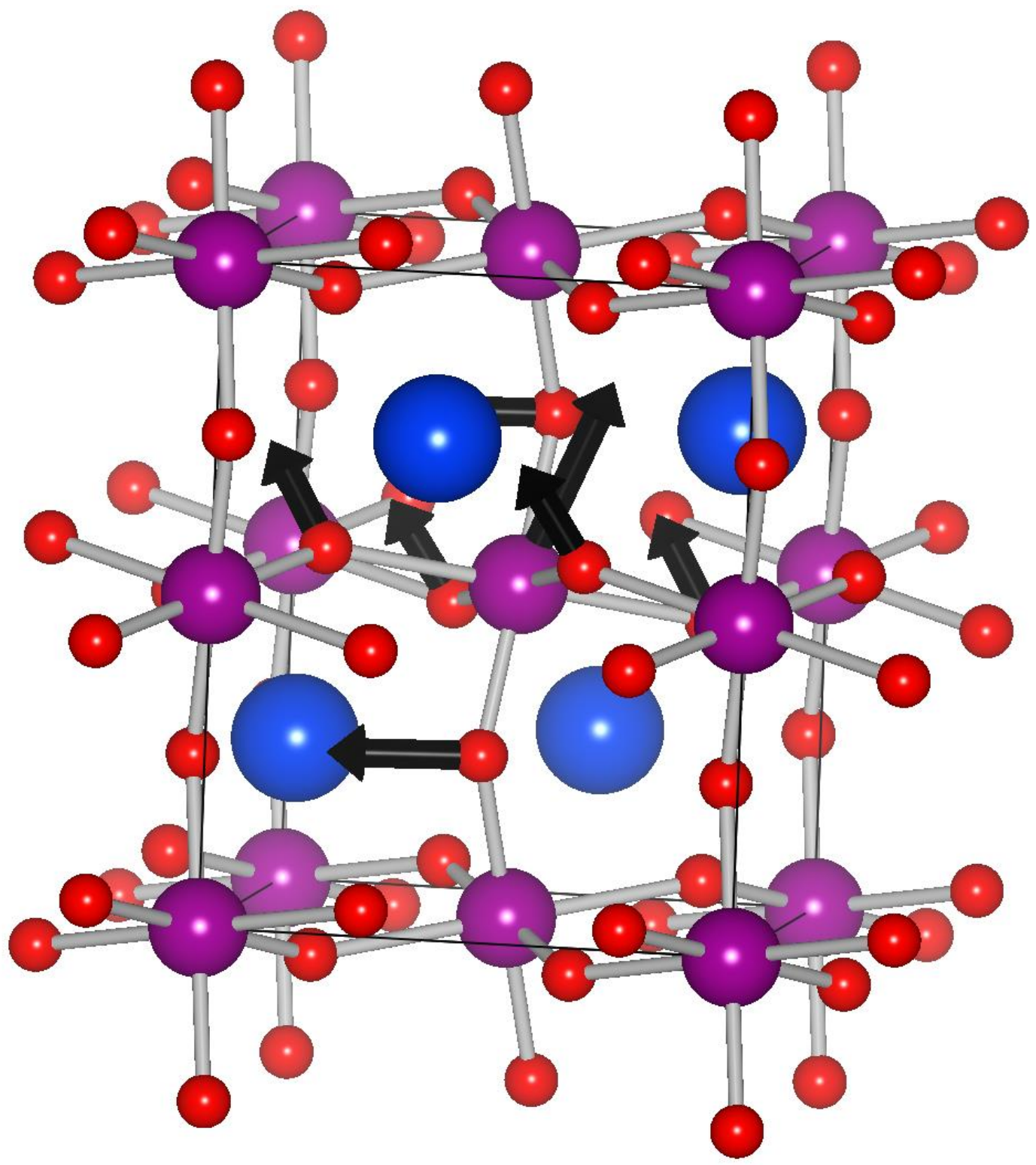

Fig. 12.  Wu *et al*.

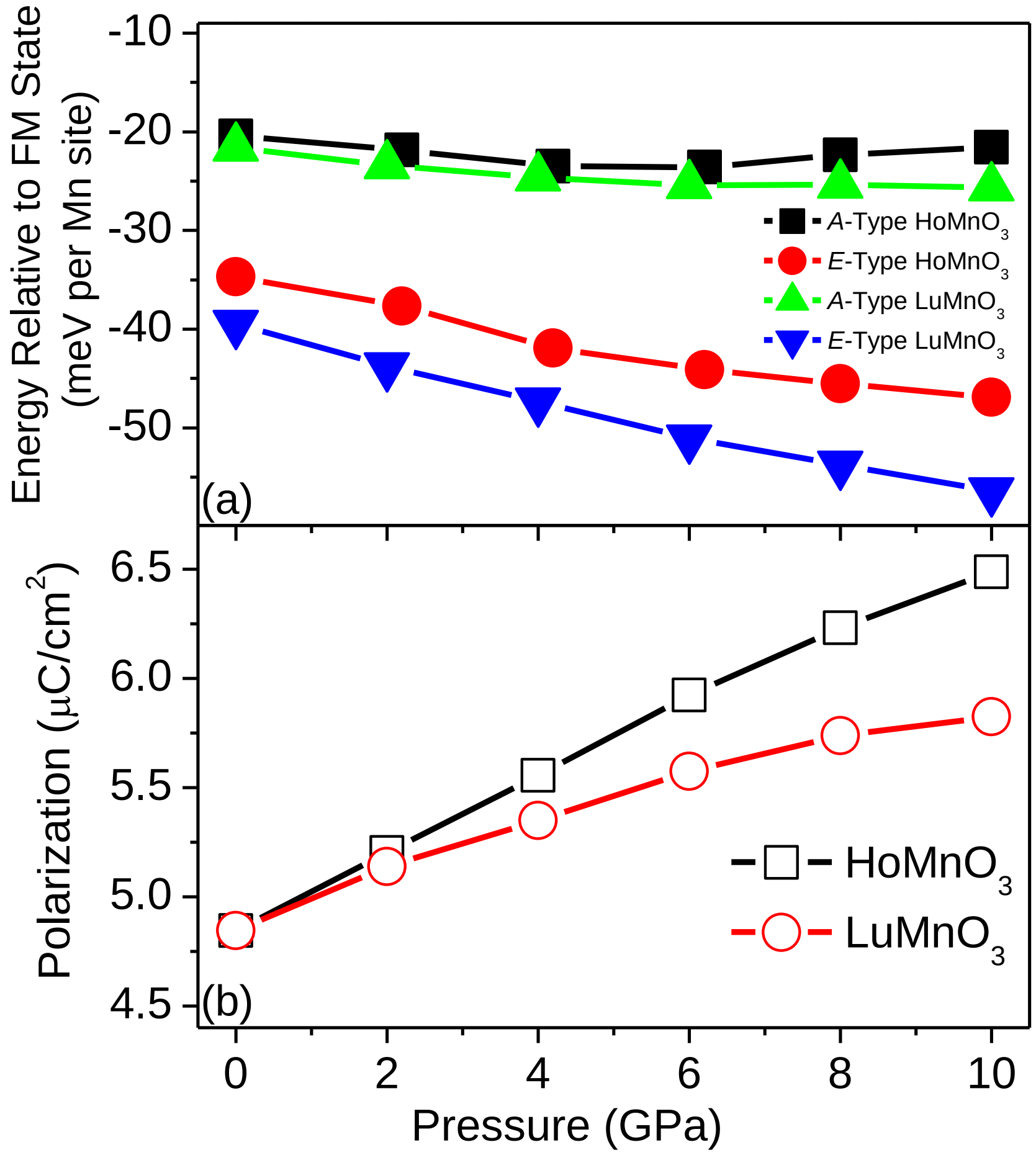

## References

1. K. F. Wang, J. M. Liu and Z. F. Ren, Advances in Physics **58**, 321 (2009).
2. M. Bibes and A. Barthelemy, Nature Materials **7**, 425 (2008).
3. W. Prellier, M. P. Singh and P. Murugavel, Journal of Physics: Condensed Matter **17**, R803 (2005).
4. V. E. Wood, A. E. Austin, E. W. Collings and K. C. Brog, Journal of Physics and Chemistry of Solids **34**, 859 (1973).
5. S. Quezel, J. Rossat-Mignod and E. F. Bertaut, Solid State Communications **14**, 941 (1974).
6. T. Kimura, T. Goto, H. Shintani, K. Ishizaka, T. Arima and Y. Tokura, Nature **426**, 55 (2003).
7. T. Goto, T. Kimura, G. Lawes, A. P. Ramirez and Y. Tokura, Physical Review Letters **92**, 257201 (2004).
8. T. Kimura, S. Ishihara, H. Shintani, T. Arima, K. T. Takahashi, K. Ishizaka and Y. Tokura, Physical Review B **68**, 060403 (2003).
9. M. Tachibana, T. Shimoyama, H. Kawaji, T. Atake and E. Takayama-Muromachi, Physical Review B **75**, 144425 (2007).
10. M. Kenzelmann, A. B. Harris, S. Jonas, C. Broholm, J. Schefer, S. B. Kim, C. L. Zhang, S. W. Cheong, O. P. Vajk and J. W. Lynn, Physical Review Letters **95**, 087206 (2005).
11. T. Arima, A. Tokunaga, T. Goto, H. Kimura, Y. Noda and Y. Tokura, Physical Review Letters **96**, 097202 (2006).
12. I. A. Sergienko, C. Şen and E. Dagotto, Physical Review Letters **97**, 227204 (2006).
13. I. A. Sergienko, C. Şen and E. Dagotto, Physical Review Letters **97**, 227204 (2006).
14. S. Picozzi, K. Yamauchi, B. Sanyal, I. A. Sergienko and E. Dagotto, Physical Review Letters **99**, 227201 (2007).
15. A. Munoz, M. T. Casais, J. A. Alonso, M. J. Martinez-Lope, J. L. Martinez and M. T. Fernandez-Diaz, Inorganic Chemistry **40**, 1020 (2001).
16. B. Lorenz, Y. Q. Wang, Y. Y. Sun and C. W. Chu, Physical Review B **70**, 212412 (2004).
17. J. S. Zhou, J. B. Goodenough, J. M. Gallardo-Amores, E. Morán, M. A. Alario-Franco and R. Caudillo, Physical Review B **74**, 014422 (2006).
18. B. Lorenz, Y.-Q. Wang and C.-W. Chu, Physical Review B **76**, 104405 (2007).
19. N. Lee, Y. J. Choi, M. Ramazanoglu, W. Ratcliff, II, V. Kiryukhin and S. W. Cheong, Phys Rev B **84**, 020101 (2011).
20. D. Iuşan, K. Yamauchi, P. Barone, B. Sanyal, O. Eriksson, G. Profeta and S. Picozzi, Physical Review B **87**, 014403 (2013).

21. S. M. Feng, Y. S. Chai, J. L. Zhu, N. Manivannan, Y. S. Oh, L. J. Wang, Y. S. Yang, C. Q. Jin and K. Kee Hoon, New Journal of Physics **12**, 073006 (2010).
22. H. Okamoto, N. Imamura, B. C. Hauback, A. Karppinen, H. Yamauchi and H. Fjevag, Solid State Communications **146**, 152 (2008).
23. K. Uusi-Esko, J. Malm, N. Imamura, H. Yamauchi and M. Karppinen, Mater Chem Phys **112**, 1029 (2008).
24. M. Garganourakis, Y. Bodenthin, R. A. de Souza, V. Scagnoli, A. Dönni, M. Tachibana, H. Kitazawa, E. Takayama-Muromachi and U. Staub, Physical Review B **86**, 054425 (2012).
25. Y. S. Chai, Y. S. Oh, L. J. Wang, N. Manivannan, S. M. Feng, Y. S. Yang, L. Q. Yan, C. Q. Jin and K. H. Kim, Physical Review B **85**, 184406 (2012).
26. S. Ishiwata, Y. Kaneko, Y. Tokunaga, Y. Taguchi, T. Arima and Y. Tokura, Physical Review B **81**, 100411 (2010).
27. I. Loa, P. Adler, A. Grzechnik, K. Syassen, U. Schwarz, M. Hanfland, G. K. Rozenberg, P. Gorodetsky and M. P. Pasternak, Physical Review Letters **87**, 125501 (2001).
28. J. M. Chen, T. L. Chou, J. M. Lee, S. A. Chen, T. S. Chan, T. H. Chen, K. T. Lu, W. T. Chuang, H. S. Sheu, S. W. Chen, C. M. Lin, N. Hiraoka, H. Ishii, K. D. Tsuei and T. J. Yang, Physical Review B **79**, 165110 (2009).
29. J. M. Chen, J. M. Lee, T. L. Chou, S. A. Chen, S. W. Huang, H. T. Jeng, K. T. Lu, T. H. Chen, Y. C. Liang, S. W. Chen, W. T. Chuang, H. S. Sheu, N. Hiraoka, H. Ishii, K. D. Tsuei, E. Huang, C. M. Lin and T. J. Yang, The Journal of Chemical Physics **133**, 154510 (2010).
30. J. Oliveira, J. Agostinho Moreira, A. Almeida, V. H. Rodrigues, M. M. R. Costa, P. B. Tavares, P. Bouvier, M. Guennou and J. Kreisel, Physical Review B **85**, 052101 (2012).
31. A. P. Hammersley, S. O. Svensson and A. Thompson, Nuclear Instruments and Methods in Physics Research Section A **346**, 312 (1994).
32. F. D. Murnaghan, Proceedings of the National Academy of Sciences **30**, 244 (1944).
33. C. Meneghini, D. Levy, S. Mobilio, M. Ortolani, M. Nuñez-Reguero, A. Kumar and D. D. Sarma, Physical Review B **65**, 012111 (2001).
34. T. Mizokawa, D. I. Khomskii and G. A. Sawatzky, Physical Review B **60**, 7309-7313 (1999).
35. A. Arulraj and C. N. R. Rao, Journal of Solid State Chemistry **145**, 557-563 (1999).
36. I. S. Smirnova, Physica B: Condensed Matter **262**, 247-261 (1999).
37. I. Fedorov, J. Lorenzana, P. Dore, G. De Marzi, P. Maselli, P. Calvani, S. W. Cheong, S. Koval and R. Migoni, Physical Review B **60**, 11875-11878 (1999).
38. J. T. Last, Physical Review **105**, 1740-1750 (1957).
39. F. Gao, X. L. Wang, M. M. Farhoudi and R. A. Lewis, Magnetics, IEEE Transactions on **41**, 2763-2765 (2005).
40. R. Sopracase, G. Gruener, E. Olive and J.-C. Soret, Physica B: Condensed Matter **405**, 45-52 (2010).
41. R. Schleck, R. L. Moreira, H. Sakata and R. P. S. M. Lobo, Physical Review B **82**, 144309 (2010).

42. R. Choithrani, M. N. Rao, S. L. Chaplot, N. K. Gaur and R. K. Singh, Journal of Magnetism and Magnetic Materials **323**, 1627-1635 (2011).
43. S. Issing, "Correlation Between Lattice Dynamics and Magnetism in Multiferroic Manganites", Ph. D. Thesis (2011).

43. (a) K. Parlinski, Z. Q. Li, and Y. Kawazoe, Phys. Rev. Lett. 78, 4063 (1997)

(b) A. Togo, F. Oba, and I. Tanaka, Phys. Rev. B, **78**, 134106 (2008).